\documentstyle[prl,aps,epsfig,epsf]{revtex}
\begin{document}
\draft
\title{Decaying magnetohydrodynamics: effects of initial conditions}
\author{Abhik Basu\cite{bypur}}
\address{Centre for Condensed Matter Theory,
Department of Physics, Indian Institute of Science, Bangalore, 560012,
INDIA}
\maketitle
\sloppy
\begin{abstract}
We study the effects of random homogenous and isotropic initial conditions on
decaying Magnetohydrodynamics (MHD). We show that for an initial
distribution of velocity and magnetic field fluctuations,
appropriately defined structure functions
 decay as power law in time. We also show that
for a suitable choice of initial cross-correlations between velocity
and magnetic fields even order structure functions acquire anomalous
scaling in time where as scaling exponents of the odd order structure functions
remain unchanged. We discuss our results in the context of fully developed
MHD turbulence.
\end{abstract}


\section{Introduction}
In the absence of external forcing the fluid cannot maintain its steady state
because of the continous dissipation of energy. The decay of kinetic
energy in incompressible fluid turbulence has been explored in several studies
(see, e.g., \cite{4smith,4bellot,4low}). These studies, principally numerical,
 indicate that
kinetic energy decays as $\sim (t-t_o)^{-\eta}$, where $t_o$ is the initial
time, i.e. the time when the initial forcing is switched off.
The value of $\eta$ seems to depend on  the initial state. 
Similarly, recent numerical studies of decaying MHD turbulence 
\cite{4low,4hossain,4pouquet,bisprl}
 also suggest that the total energy decays as power of time.
The phenomena
of decaying MHD turbulence is observed in many different situations.
Star-forming clouds are astrophysical examples \cite{4low}.

\subsection{Decay of MHD turbulence in a shell model}
In Ref.\cite{abprl} we described a new shell model for MHD turbulence,
which has all the conservation laws in the inviscid, unforced limit,
has the right symmetrires and reduces to the well-known GOY shell
model for fluid turbulence in absence of any magnetic field. 
In this paper we use our shell model to study decaying MHD turbulence.
We begin by solving this set of equations with the external forces 
and by using the numerical method described there. We evolve the shell velocities
$v_n$ and magnetic fields $b_n$ till we achieve the nonequilibrium, statistical
steady state whose properties are described in Ref.\cite{abprl}. 
Once we attain this
state, say at a time $t=t_{ss}$, we use the last set of shell velocities and
magnetic fields, i.e., $v_n(t_{ss})$ and $b_n(t_{ss})$, as the initial condition
for our shell-model study of decaying turbulence in which we solve our 
shell-model equations   
but with all the forcing terms set equal to zero. We then use the 
resulting time series for $v_n$ and $b_n$ to determine the total energy 
$E(t)\equiv \sum_{\bf k} \langle|{\bf v}({\bf k},t)|^2\rangle+
\langle|{\bf b}({\bf k},t)|^2\rangle$ and the total crosshelicity 
$H^c(t)\equiv \langle|{\bf v}({\bf k},t).{\bf b}({\bf k},t)|\rangle$ 
where the origin of time is taken to be zero. We find that $E(t)
\sim H^c(t)\sim t^{-p}$, with 
$p\simeq 1.2$; this is illustrated in the log-log plot of Fig.1. 
As we have noted
above, it is relevant for initial conditions that have the correct statistical
properties of homogenous, isotropic MHD turbulence which are well decribed by our
shell model.

\begin{figure}[htb]
\epsfxsize=5.5cm
\epsfysize=5.5cm
\vspace{-0.4cm}
\centerline{\epsfbox{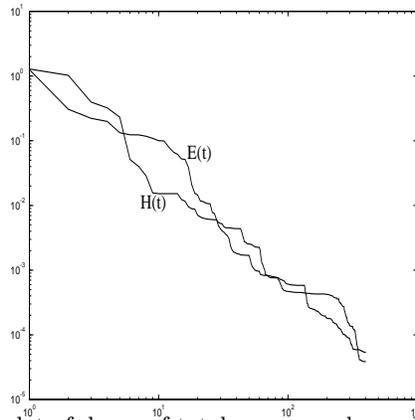}}
\vspace{-0.2cm}
\caption{A log-log plot of decay of total energy and crosshelicity versus time.}
\end{figure}

In a recent paper \cite{4peri} the effects of random initial
conditions on the decaying solutions of the unforced Navier-Stokes equation
with hyperviscosity have been discussed: It has been shown that, for
an initially Gaussian velocity distribution,
equal-time two point correlation
functions like $<u(x,t)u(x,t)>$ decay as $\sim t^{-1}$. It has
also been shown that, for this particular choice of initial conditions,
the theory remains finite (in a field-theoretic sense)
just by renormalisation of the two-point vertex
functions.  However, it is more relevant to explore the statitical properties
of decaying fluid turbulence with ordinary viscosity. It is also important
to find out how the introduction of magnetic fields influences this decay.
In this paper we study decaying MHD with ordinary viscosity; 
we work with both the
$3d$MHD equations and a $d$-dimensional generalisation of the $1d$MHD model
introduced in Ref.\cite{epjb}. We show that both these models give similar results. 
In particular, we show that, in both these models, appropriately
defined structure functions decay with a power law in time; this is
modulated by a scaling
function whose form depends on the initial state. In 
Section.II, we describe the equations we work with and the
intial conditions we use.
For most of the calculations in this Chapter,
we choose initial correlations which are analytic in the low-wavenumber
limit. We choose ordinary viscosities 
for both velocity and magnetic fields. We calculate the {\em dynamic
exponent} $z$ and show that it is unaffected by the presence of
the nonlinearities, i.e., it remains 2. We next show that if there is no 
 initial cross-correlation between velocity 
and magnetic fields, correlation functions defined at the same spatial point
decay with power laws in time and, consequently,
two point structure functions also decay likewise.
Also,  multipoint correlation functions (and structure functions)
exhibit similar behaviour, with simple (and not anomalous) 
scaling in time.  In other words, the exponents for high-order
structure functions are simply related to
those of two-point functions. Another interesting feature is that
there is no change in the exponents
with the introduction of the magnetic field (i.e. the exponents are the same
for fluid and MHD turbulence); only the amplitudes of the
correlation or structure functions change. 
In the Section III we show that in the
presence of suitable cross correlations higher order correlations (and
consequently structure functions) decay
more slowly than when there is no cross correlation, i.e.,
they display anomalous scaling. 
In Section IV we extend and discuss our results in the context of
the decay of fully developed MHD turbulence. In Section V we conclude  
and compare results obtained from both $3d$MHD and our simplified model.

\section{Decaying MHD turbulence: the case of zero initial
cross-correlation of the velocity and magnetic fields}

\subsection{Construction of a simplified model}
The dynamical equations which govern the time evolutions of the velocity and
magnetic fields in a magnetised fluid are given by the equations of
Magnetohydrodynamics, namely,
\begin{eqnarray}
{\partial {\bf v}\over\partial t}+{\bf (v.\nabla)v}&=&-{\nabla p\over\rho}+
{\bf (\nabla \times b) \times b\over\ 4\pi}+\nu{\bf \nabla^2 v}+{\bf f_v},\\
{\partial {\bf b}\over\partial t}&=&\nabla \times {\bf (v\times b)}
+\mu{\bf \nabla^2 b}+{\bf f_b},
\end{eqnarray}
where $\bf v$ and $\bf b$ are the velocity and the magnetic fields,
$\nu$ and $\mu$ are fluid and magnetic viscosities, $p$ and $\rho$
represent pressure and density, and $\bf f_v$ and $\bf f_b$ are external
forces.
We also impose the incompressibility condition $\nabla \bf .v=0$ and the
divergence-free condition on the magnetic field $\bf \nabla .b=0$.
In the absence of magnetic fields the MHD equations reduce 
to usual Navier-Stokes equation for fluid turbulence. 

We now briefly present the ideas that go into the construction of the model,
which has already been discussed in details in \cite{epjb}. Since this model
is an extension of Burgers equation for fluid dynamics to MHD, it should
have the following properties:
\begin{itemize}
\item Both the equations should be Galilean invariant.

\item The equations should be invariant under $t\rightarrow t;
\bf x\rightarrow -x$, $\bf v\rightarrow -v$ and $\bf b\rightarrow b,$
i.e. the velocity field is odd parity and the magnetic field is even parity.

\item The fields are {\em irrotational} vector fields in $d$-dimension (
similar to Burgers velocity field).

\item The equations in the ideal limit i.e. in absence of any viscosity and
any external forcing should conserve all the scalar conserved quantities of
$3d$MHD namely the total energy and the cross-helicity. The third
conserved quantity i.e. the magnetic helicity is zero for irrotational fields
.
\end{itemize}                                                          
The above considerations lead to the following model equations
\begin{equation}
{\bf \partial v\over\partial t}+{1\over\ 2}\nabla v^2 +{1\over\ 2}\nabla b^2
=\nu\nabla ^2 {\bf v} + {\bf f_v},
\label{mhd1}
\end{equation}
\begin{equation}
{\bf \partial b\over\partial t}+\nabla ({\bf v.b})=\mu\nabla^2 {\bf b}+
{\bf f_b},
\label{mhd2}
\end{equation}
whose relation to the $3d$MHD equations is the same as the relation of the
Burgers equation to the Navier-Stokes equation.
Here $\bf v$ and $\bf b$ are the velocity and magnetic fields with
magnitudes $v$ and $b$ respectively. These fields
are $irrotational$, i.e., $\bf \nabla\times v=0$ and $\bf \nabla \times b=0$. 
These equations together conserve the total energy $1/2 \int(v^2+b^2),d^dx$ and 
crosshelicity $\bf \int v.b\,d^dx$.  

\subsection{Choice of initial conditions}
We have $<u_i(k,t=0)>=<b_i(k,t=0)> =0$ by homogeneity and isotropy.
We choose the initial correlations to have the simple analytic forms
\begin{eqnarray}
<v_i(k,t=0)v_j(k^{\prime},t=0)>=D_{uu}k_ik_j\delta(k+k^{\prime}), \\
<b_i(k,t=0)b_j(k^{\prime},t=0)>=D_{bb}k_ik_j\delta(k+k^{\prime}), 
\end{eqnarray}
and
\begin{equation}
<v_i(k,t=0)_ib_j(k^{\prime},t=0)>=D_{ij}(k)\delta(k+k^{\prime})
\end{equation}

In $k$-space the $3d$MHD equations are 
\begin{eqnarray}
\partial_t v_i({\bf p})&+&iM_{ijk}({\bf p})\sum_q v_j({\bf q})v_k ({\bf p-q})= 
iM_{ijl}({\bf p})\sum_q b_j({\bf q})b_l ({\bf p-q})/ 4\pi\rho  \nonumber \\
&-&\nu p^2 v_i({\bf p}) +f_i({\bf p})\\
\label{3dmhdu}
\partial_t b_i({\bf p})&=&-
(i{\bf p}\times \sum_q v({\bf q})\times b({\bf p-q}) )_i
-\mu p^2 b_i({\bf p}) +g_i({\bf p})
\label{3dmhdb}
\end{eqnarray}
The quantities $M_{ijl}({\bf k})=P_{ij}k_l+P_{il}k_j$ appear to include 
the incompressibility, $P_{ij}(k)=(\delta_{ij}-{k_ik_j\over k^2}$) 
is the transverse projection operator and wavevector arguments $p,q,k$
indicate spatial Fourier transforms.
The initial conditions we use for $3d$MHD are
\begin{eqnarray}
<u_i(k,t=0)u_j(k^{\prime},t=0)>&=&
D_{uu}k^2P_{ij}({\bf k})\delta(k+k^{\prime}), \\
<b_i(k,t=0)b_j(k^{\prime},t=0)>&=&
D_{bb}k^2P_{ij}({\bf k})\delta(k+k^{\prime}), \\
<u_i(k,t=0)b_j(k^{\prime},t=0)>&=&D_{ij}(k)\delta(k+k^{\prime})
\end{eqnarray}
which ensure the divergence-free conditons, namely,
\begin{eqnarray}
k_i<u_i({\bf k})u_j {(\bf -k)}>&=&0 \\
k_i<b_i({\bf k})b_j {(\bf -k)}>&=&0.
\end{eqnarray}
We discuss the structure of $D_{ij}$ later.
For the time being, we choose $D_{ij}=0$. These simple initial conditions,
which are analytic in the $k\rightarrow 0$ limit,
are enough to elcidate the issues raised in this Chapter.

\subsection{Calculation of effective correlation and structure 
functions within a one-loop perturbation theory: temporal scaling}

The formal solutions of the Eqs. (3) and (4) are given by
\begin{eqnarray}
u_l ({\bf k},t)&=&e^{-\nu k^2 t} u_l({\bf k},0)-{i\over\ 2} k_l \int_0 ^t
dt^{\prime}\int d^dp
\exp[-\nu(t-t^{\prime})k^2]\exp\{-\nu t^{\prime}
[p^2+(k-p)^2]\}\nonumber \\&&[u_m ({\bf p},0)u_m ({\bf k-p},0)
+ \exp\{-\mu t^{\prime}[p^2+(k-p)^2]\}\times\nonumber \\
&&b_m({\bf p},0)b_m({\bf k-p},0)] +...
\label{mu1}
\end{eqnarray}
\begin{eqnarray}
b_l ({\bf k},t)&=&e^{-\mu k^2 t}b_l({\bf k},0)-
ik_l\int_0 ^t dt^{\prime}\int d^dp
\exp[-\mu (t-t^{\prime})k^2]\exp\{-t^{\prime}[\nu p^2+\mu (k-p)^2]\}\nonumber \\
&&[u_m({\bf p},0)b_m({\bf k-p},0)]+...
\label{mb1}
\end{eqnarray}
where ellipsis refer to higher order terms
Similarly the formal solutions of the $3d$MHD equations (\ref{3dmhdu}) and
(\ref{3dmhdb}) are given by:
\begin{eqnarray}
u_l ({\bf k},t)&=&e^{-\nu k^2 t} u_l({\bf k},0)-{i\over\ 2} M_{ljs}\int_0 ^t
dt^{\prime}\int d^dp
\exp[-\nu(t-t^{\prime})k^2]exp\{-\nu t^{\prime}
[p^2+(k-p)^2]\}\nonumber \\&&[u_j ({\bf p},0)u_s ({\bf k-p},0)
\nonumber \\ &&- \exp\{-\mu t^{\prime}[p^2+(k-p)^2]\}b_j
({\bf p},0)b_s({\bf k-p},0)]+...,
\label{3dmu1}
\end{eqnarray}
\begin{eqnarray}
b_l ({\bf k},t)&=&e^{-\mu k^2 t}b_l({\bf k},0)-
ik_m\int_0 ^t dt^{\prime}\int d^dp
\exp[-\mu (t-t^{\prime})k^2]\exp\{-t^{\prime}[\nu p^2+\mu (k-p)^2]\}
\nonumber \\&& [u_m({\bf p},0)b_l({\bf k-p},0) -u_l({\bf p},0)b_m({\bf k-p},0)]
+...,
\label{3dmb1}
\end{eqnarray}
where the ellipsis in refer to higher order terms in the series. 
We use bare perturbation theory, i.e., after truncating the series 
at a particular order, all $\bf v$ and $\bf b$ are to be replaced by their
bare values, i.e., by $G_o^u u(t=0)$ and $G_o^b b(t=0)$, respectively, 
where $G_o^u({\bf k},t)=e^{-\nu k^2 t}$ and $ G_o^b({\bf k},t)=e^{-\mu k^2 t}$. This
is similar to Ref.\cite{4fns}.
In the steady state of fluid or MHD turbulence, the {\em dynamic exponent} $z$ 
is determined by the form of the correlations  
of the driving noise and dimensionality of the space. For example, when
driven by noise correlations of the form $k^{-3}$ (in $3d$NS), one finds $z$ 
(within a 1-loop dymanic renormalisation group calculation) to be 2/3\cite
{4yakhot}. What is $z$ when we study decaying MHD turbulence? We
calculate this within a 1-loop diagramatic perturbation theory:
In the bare theory (i.e., when the nonlinear terms are absent) $z=2$.
If the one-loop corrections to the response functions (i.e., to the
viscosities) have infrared divergence at zero external frequency,
then $z<2$; otherwise $z=2$. The diagrams shown in Fig.2 contribute at the
one-loop level in renormalisation of the fluid viscosity $\nu$,
\begin{figure}
\epsfxsize=13cm
\centerline{\epsfbox{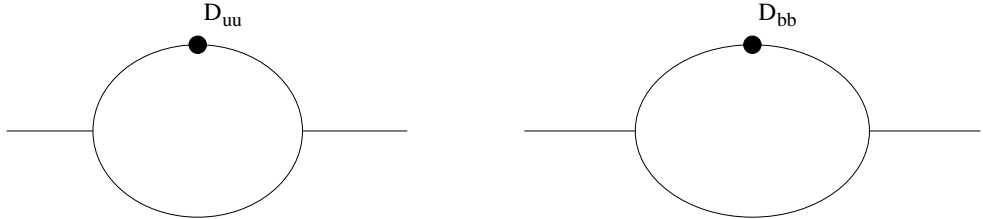}}
\caption{1-loop diagrams contributing to the renormalisation of $\nu$.}
\end{figure}
Similar diagrams renormalise $\mu$. 
The effective 1-loop fluid viscosity is given by
\begin{eqnarray}
{1\over\tilde{\nu}k^2}&=&{1\over\nu k^2}+{D_{uu}\over\nu k^2}
\int d^3 p{p_l p_m 
{\bf k.(k-p)}\over\ 2\nu {\bf k.p}}[{1\over\ 2\nu p^2 -2\nu{\bf k.p}} 
(1-{k^2\over\nu(p^2+({\bf k-p})^2}))\nonumber \\&-&{1\over\ 2\nu p^2}
(1-{k^2\over\ 2\nu p^2})]
+{D_{bb}\over\nu k^2}\int d^3 p{p_l p_m {\bf k.(k-p)}\over\ 2\mu {\bf k.p}}
[{1\over\ 2\mu p^2 }(1-{\nu k^2\over\ 2\mu p^2})\nonumber \\ &-&
{1\over\ 2\mu p^2 }(1-{\nu k^2\over\ 2\mu p^2})].
\end{eqnarray}
We see that, in the long-wavelength limit, 1-loop corrections
have no infrared divergence at $d=3$; in other words 
$z$ remains 2. In the case of the $3d$MHD equations, 
the forms of the integrals remain
unchanged, only changes appear in the coefficients of the integrals
where, in place of $p_lp_m$, appropriate projection operators 
are present. Thus again there
is no diagramatic correction to viscosities in the long-wavelenght limit in 
$3d$MHD. Hence the exponent $z$
is unaffected by the nonlinear terms. Therefore, without any loss
of generality, we can put $\nu=\mu=1$.

We now calculate the effective $D_{uu}$ and $D_{bb}$ for $d=3$
at the one-loop level.
The diagrams shown in Fig.3 contribute at the 1-loop level to $D_{uu}$,
These contributions are proportional to
\begin{figure}
{\epsfxsize=13cm
\centerline{\epsfbox{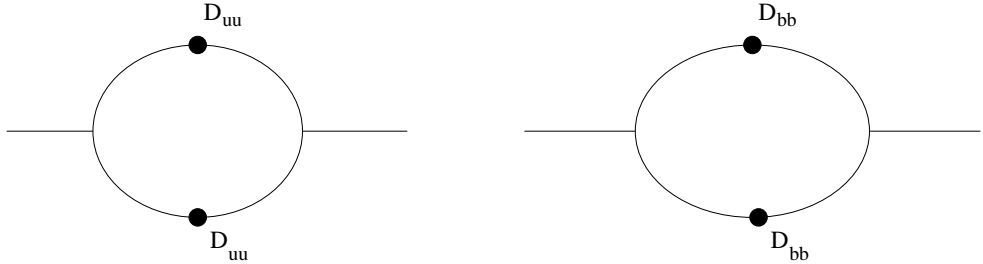}}}
\caption{1-loop diagrams contributing to the renormalisation of
$D_{uu}$ and $D_{bb}$ respectively.}
\end{figure}
\begin{equation}
k_i k_j (D_{uu}^2+D_{bb}^2)\int d^3 p {p^2 p^2 (p^2-k^2)^2\over k^4 p^8}
\sim k_i k_j (D_{uu}^2+D_{bb}^2)\int^{\Lambda} d^3p,
\end{equation}
which does not have any infrared divergence in {\em any dimension}. 
The one-loop diagram contributing to $D_{bb}$ has the same form.
For $3d$MHD the forms of the one-loop integrals do not change; only 
the coefficients have appropriate factors of projection
operators instead of $k_ik_j$. On purely dimensional grounds for $d=3$,
\begin{equation}
\langle u_i({\bf x},t)u_j({\bf x},t) \rangle \sim \int d^3 k
\langle u_i({\bf k},t)u_j(-{\bf k},t)
\rangle \sim D_{uu}\int d^3 k e^{-\nu k^2 t} e^{-\nu k^2 t} k_i k_j 
\sim D_{uu}t^{-5\over\ 2}\delta_{ij} 
\end{equation}
and
\begin{equation}
\langle b_i({\bf x},t)b_j({\bf x},t) \rangle \sim \int d^3 k
\langle b_i({\bf k},t)b_j(-{\bf k},t)
\rangle D_{bb}\sim \int d^3 k e^{-\nu k^2 t} e^{-\nu k^2 t} k_i k_j
\sim D_{bb}t^{-5\over\ 2}\delta_{ij}.
\end{equation}
Similar quantities obtained from $3d$MHD also show the same exponents but with
different amplitudes \footnote{This is easy to understand: Calculation of the
exponents through the one-loop method depends upon the power-counting
of the one-loop integrals which are same for both the $3d$MHD equations and
our simplified model as the projection operator $P_{ij}$ that appears in
the $3d$MHD equations is dimensionless and the nonlinearities in both
the equations do not renormalise due to the Galilean invariance of our
simplified model and $3d$MHD equations (which in turn keeps the power-counting
structure same). Amplitudes of course depend upon the
specific models. Similarly, energy spectra obtained in the steady-state 
from different models which give same power-counting in the one-loop integrals,
when driven by stochastic forces with correlations having same scaling, 
have same scaling propertes, see, for e.g. Refs.\cite{epjb,4yakhot}.}. 
Note that the temporal dependences can be scaled out
of the momentum integrals easily as the lower and upper limits of the 
momentum integrals can be extended to 0 and $\infty$, respectively.

\begin{figure}
\epsfxsize=12cm
\epsfysize=7cm
\centerline{\epsffile{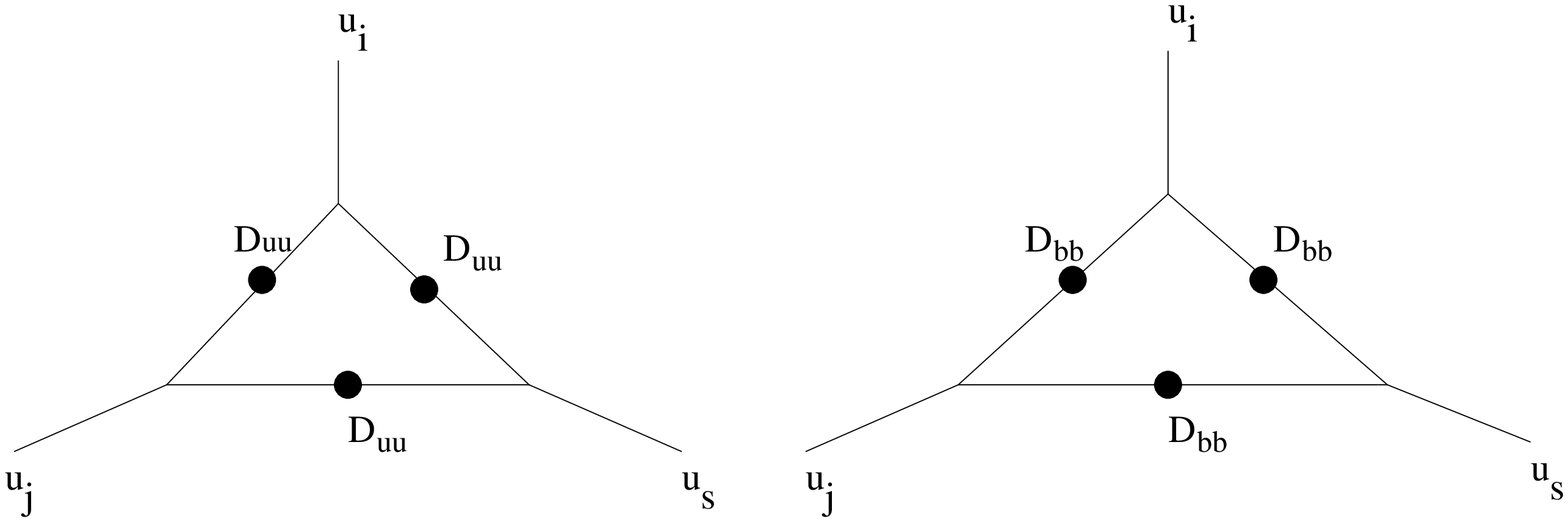}}
\caption{\small One-loop diagrams contributing to 
$\langle u_({\bf k_1},t)u_j({\bf k_2},t)u_s({\bf k_3},t)\rangle$.}
\label{duuu}
\end{figure}

It is easy to see that all multi-point correlation functions also
do not exhibit any infrared divergence in three dimensions: Consider, e.g.,
3-point functions. Even though $\langle u_({\bf k_1},t)
u_j({\bf k_2},t)u_s({\bf k_3},t)\rangle 
= 0$ at $t=0$, it becomes non-zero at $t>0$ as can be easily seen at the
one-loop level. 
$\langle u_({\bf k_1},t)u_j({\bf k_2},t)u_s({\bf k_3},t)\rangle$ has
a tree-level contribution and next a one-loop contribution. 
The tree-level part being proportional to 
$\langle u_i({\bf k_1},0)u_j({\bf k_2},0)
u_s({\bf k_3},0)\rangle$ is zero because we have chosen a Gaussian distribution
for the initial conditions.
The one-loop diagrams are shown in Fig.(\ref{duuu}).
 The 1-loop contribution is however nonzero. It has the form
\begin{equation}
\langle u_({\bf k_1},t)u_j({\bf k_2},t)u_s({\bf k_3},t)\rangle
\sim  \exp[\nu(-k_1^2-k_2^2-({\bf k_1+k_2})^2)t] k_{1i}k_{2j}(-k_{1s}-k_{2s})
D_{uuu},
\end{equation}
where $D_{uuu}$ is the value of the one-loop integral in the 
limit $k_1,k_2\rightarrow 0$, i.e.,
\begin{equation}
D_{uuu}\sim\int d^dq (D_{uu}^3+D_{bb}^3)
\end{equation}
which is infrared (IR) finite. We can interpret this
one-loop result by saying that the nonlinearities in Eqs.(\ref{mhd1}) and
(\ref{mhd2}) are absent but there are nonzero
initial three-point correlations at $t=0$, i.e., the effective linear system
behaves as if
\begin{eqnarray}
<u_i({\bf k_1},0)u_j({\bf k_2},0)u_s({\bf k_3},0)>&
\equiv&k_{1i}k_{2j}k_{3s}D_{uuu}
\delta({\bf k_1+k_2+k_3}), \\
<b_i({\bf k_1},0)b_j({\bf k_2},0)b_s({\bf k_3},0)>&
\equiv&k_{1i}k_{2j}k_{3s}D_{bbb}
\delta({\bf k_1+k_2+k_3}). 
\end{eqnarray}
Hence
\begin{eqnarray}
&&<u_i({\bf x_1},t)u_j({\bf x_2},t)u_s({\bf x_3},t)>\nonumber \\  &\equiv&
\int d^3k_1 d^3k_2 \exp[i{\bf k_1.(x_1-x_2)}+{\bf k_2.(x_1-x_3)}]
\langle u_i({\bf k_1},t)u_j({\bf k_2},t)u_s({\bf -k_1-k_2},t)> \nonumber \\
&=&\int d^3k_1 d^3k_2 exp(-k_1 ^2-k_2^2 -({\bf k_1+k_2})^2)t
\exp[i{\bf k_1.(x_1-x_2)+k_2.(x_1-x_3)}] 
\nonumber \\&&k_{1i}k_{2j}k_{3s}D_{uuu}
\sim t^{-9/2}A_{ijs}({{\bf x_1-x_2}\over\sqrt t},{{\bf x_1-x_3}\over\sqrt t}).
\end{eqnarray}
The last step can be easily obtained by using the dimensionless variables
$y_1=k_1^2t, y_2=k_2^2t$.

\begin{figure}
\epsfxsize=14cm
\centerline{\epsffile{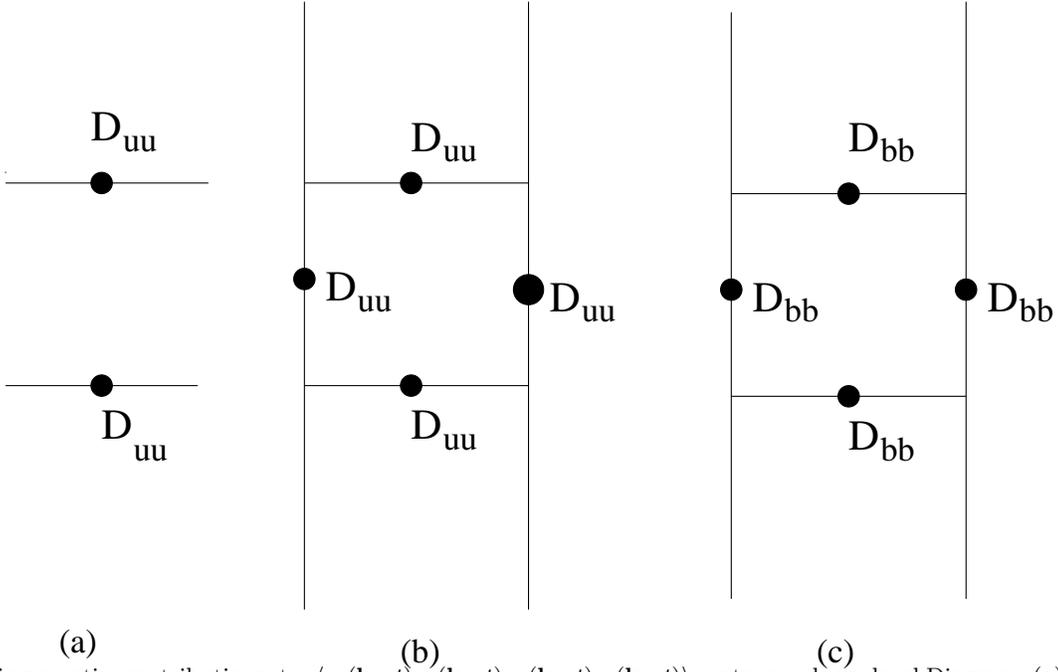}}
\caption{\small Diagramatic contributions to
$\langle u_i({\bf k_1},t)u_j({\bf k_2},t)u_s({\bf k_3},t)u_l
({\bf k_4},t)\rangle$ upto
one-loop level.Diagram (a) is the bare four-point function due to the initial
two-point correlations, diagrams (b) and (c) are one-loop contributions.}
\label{duuuu}
\end{figure}

Next we consider the four-point functions: We concentrate on
$\langle u_i({\bf x_1},t)u_j({\bf x_2},t)u_s({\bf x_3},t)u_l({\bf x_4},t)
\rangle$; the
diagramatic representations up to the one-loop contributions 
are given in Fig.(\ref{duuuu}).
The one-loop integrals in the long-wavelength limit, are
\begin{equation}
D_{uuuu}\sim\int d^3q({D_{uu}^4\over\nu^4}+{D_{bb}^4\over\mu^4}).
\end{equation}
These one-loop contributions can be interpreted as in the previous case of
three-point functions i.e. as if there were initial four-point correlations
in absence of the non-linearities:
\begin{equation}
\langle u_i({\bf k_1},0)u_j({\bf k_2},0)u_s({\bf k_3},0)u_l({\bf k_4},0)\rangle
\equiv D_{uuuu}k_{1i}k_{2j}k_{3s}k_{4l} \delta({\bf k_1+k_2+k_3+k_4}).
\end{equation}
Similar results can be easily obtained for other four-point functions
involving velocity and magnetic fields.
Hence, 
\begin{eqnarray}
&&<u_i({\bf x_1},t)u_j({\bf x_2},t)u_s({\bf x_3},t)u_l({\bf x_4},t)>
\nonumber \\ &&\sim D_{uu}^2 \int d^3 k_1 d^3 k_2 k_{1i}k_{1j}k_{2s}k_{2l}
\exp[-2\nu (k_1^2 +k_2^2)t] \exp[i{\bf k_1.(x_1-x_2)+k_2.(x_3-x_4)}]
\nonumber \\ &+&
D_{uuuu}\int d^3k_1d^3k_2d^3k_3k_{1i}k_{2j}k_{3s}(-k_{1l}-k_{2l}-k_{3l})
\exp[-\nu\{k_1^2+k^2+k_3^2+(k_1^2+k_2^2+k_3^2)\}t]
\nonumber \\&&\exp[i{\bf k_1.(x_1-x_2)+k_2(x_2-x_3)+k_3(x_3-x_4)}]\nonumber \\
&\sim& D_{uu}^2t^{-5}+D_{uuuu}t^{-13/2}. 
\label{4u}
\end{eqnarray}
In the infinite-Reynolds-number limit, i.e., in the limit $\nu,\mu
\rightarrow 0$, $D_{uuuu}\rightarrow \infty$ and hence
the second term in the last line of Eq.(\ref{4u}) dominates. As in the case 
for the three-point functions four-point functions, are of the form
\begin{equation}
<u_i({\bf x_1},t)u_j({\bf x_2},t)u_s({\bf x_3},t)u_l({\bf x_4},t)>
\sim t^{-13/2}A_{ijsl}({{\bf x_1-x_2}\over\sqrt t},
{{\bf x_2-x_3}\over\sqrt t},{{\bf x_3-x_1}\over\sqrt t}).
\end{equation}
Here $A_{ijs}$ and $A_{ijsl}$ are third and fourth
rank tensors respectively, and $<u_i({\bf x},t)^4>\sim t^{-13/2}$. 
To study decaying MHD,
we extend the definiton of equal-time structure functions as follows:
\begin{equation}
S^n_u (r,t)\equiv <|[u_i({\bf x_1},t)-u_i({\bf x_2},t)]|^n>
\end{equation}
and
\begin{equation}
S^n_b (r,t)\equiv <|[b_i({\bf x_1},t)-b_i({\bf x_2},t)]|^n>, \\
\end{equation}
where $r\equiv |\bf x_1-x_2|$ and $t$ denotes the time that has elapsed
after the forcing terms are switched off. We now look at the general forms
of these structure functions. Consider the three-point  function first:
\begin{equation}
S_u^3(r,t)\equiv \langle |[u_i({\bf x_1},t)-u_i({\bf x_2},t)]|^3\rangle.
\end{equation}
On expanding the expressions for, 
$S_u^3(r,t)$ we get the following kind of terms:
(i)$\langle u_i({\bf x_1},t)^3\rangle $ and 
(ii)$\langle u_i({\bf x_1},t)^2u_i({\bf x_2},t)\rangle$.
The first one is given by
\begin{eqnarray}
<u_i({\bf x_1},t)^3>&=&\int d^3k_1d^3k_2 k_{1i}k_{2i}(-k_{1i}-k_{2i})
\exp[-\{k_1^2-k_2^2- ({\bf k_1+k_2})^2\}t]D_{uuu}\nonumber \\
&\sim& t^{-9/2},
\end{eqnarray}
whereas the second term has the form
\begin{eqnarray}
<u_i({\bf x_1},t)^2u_i({\bf x_2},t)>&=&\int d^3k_1d^3k_2\,
k_{1i}k_{2i}(-k_{1i}-k_{2i})\exp[-\{k_1^2-k_2^2- ({\bf k_1+k_2})^2\}t]
\nonumber \\&&\exp[i{\bf k_1.(x_1-x_2)}]D_{uuu}
\sim t^{-9/2}\phi(r/\sqrt t),
\end{eqnarray}
with $r\equiv |{\bf x_1-x_2}|$. Thus, it is easy to see that $S_u^3 (r,t)$ 
has the structure
\begin{equation}
S_u^3(r,t)\equiv t^{-9/2}F_3^u (r/\sqrt t),
\end{equation}
with $F_3^u (r/\sqrt t)$ is a function with dimensionless argument $r/\sqrt t$.
We can easily extend our calculation above to see that
\begin{eqnarray}
S^n_u (r,t)\sim t^{-z_n^u} F^u_n({r\over t^{1/2}}), \\
S^n_b (r,t)\sim t^{-z_n^b} F^b_n({r\over t^{1/2}}), 
\end{eqnarray}
where $F^u_n$ and $F^b_n$ are functions 
 with dimensionless argument $r/\sqrt t$.
It is obvious that all these higher-order correlation functions {\em do not}
exhibit any {\em anomalous scaling} in time; their temporal behaviours can be
obtained by simple dimensional arguments from the behaviours of the
two-point functions.
It is easy to see that 
\begin{eqnarray}
S^n_u &\sim& t^{-(n-1)3/2-n/2} F^u_n({r\over t^{1/2}}),\\
S^n_b &\sim& t^{-(n-1)3/2-n/2} F^b_n({r\over t^{1/2}}),
\end{eqnarray}
i.e., $z_n^u=z_n^b=(n-1)3/2+n/2$, at the level of our one-loop approximation.

\section{Non-zero initial cross-correlations}
\subsection{Form of the initial cross correlations}
What happens if we make ${D}_{ij}$ non-zero? Let us first establish the
form of $D_{ij}$: Recall that
$<u_i({\bf x_1},t) b_j({\bf x_2},t)>$ has odd parity 
(since $\bf u$ is a polar vector whereas 
$\bf b$ is an axial vector).  Hence it is purely imaginary and is
an odd function of $k$. (From now on we will work only in three dimension.)
Thus we choose 
\begin{equation}
D_{ij}=iD_{ub}\epsilon_{ijp}k_p k^{\alpha},
\end{equation}
where $\epsilon_{ijp}$ is the totally antisymmetric tensor.
This is the simplest analytic form with the desired structure.
We consider the cases $\alpha=0$ and 2.

\subsection{The temporal decay of correlation and structure functions}
It is easy to see that the
value of $z$ does not change if $D_{ub}\neq 0$
as no new diagram appears. 
However, an additional diagram (Fig.4.6) contributes to the effective  
 $D_{uu}$ and $D_{bb}$ if $D_{ub}\neq 0$.
\begin{figure}[htb]
\epsfxsize=15cm
\centerline{\epsfbox{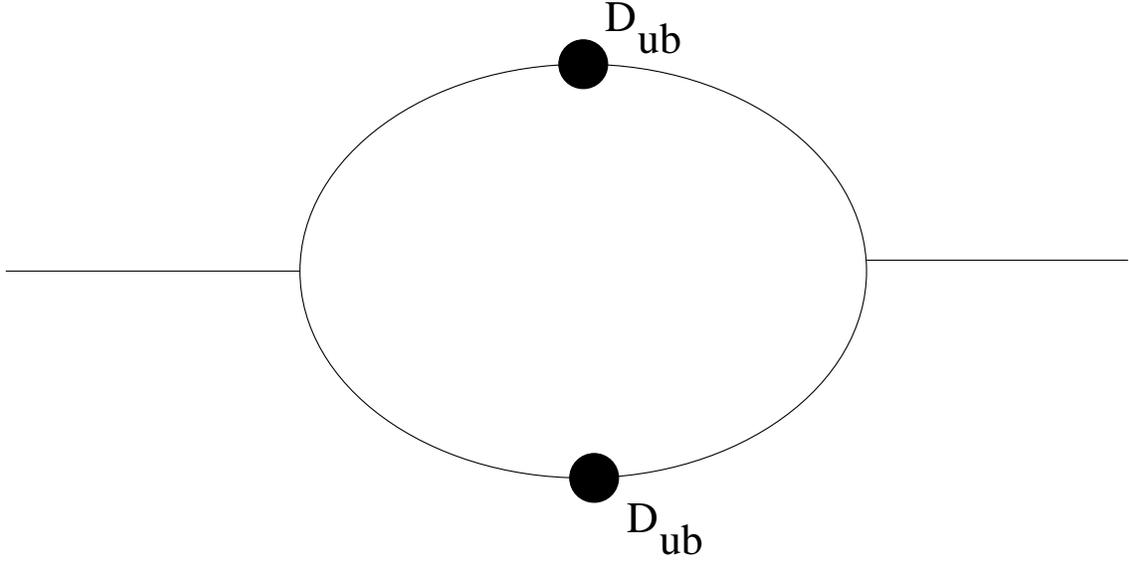}}
\caption{1-loop diagram contributing to the renormalisation of $D_{uu}$
and $D_{bb}$, which arises when $D_{ub}\neq 0$ (see text).}
\label{Fig 2}
\end{figure}
This diagram yields the following contribution 
\begin{equation}
{k_{1i}k_{2j}\over\ 16k^4}D_{ub}^2 \int {d^3q\over\ (2\pi)^3}
{2\delta_{sn} q_sq_nq^{\alpha}\over\ q^2},
\end{equation}
where we have used the identity 
\begin{equation}
\epsilon_{lms}\epsilon_{lmn}=(d-1)\delta_{sn}=2\delta_{sn}.
\end{equation}
Naive power counting shows that this contribution is
finite for $d=3$, hence $D_{uu}$ and $D_{bb}$ are not renormalised
even if $D_{ub}\neq 0$, for both $\alpha=0$ and 2. 
The same conclusion holds for $3d$MHD as
the one-loop integral remains finite; only the coefficients change because
of factors of $P_{ij}(k)$.
It is quite interesting to note that for, $D_{ub} \neq 0$, there 
is a diagramatic correction to $D_{ub}$; 
\begin{figure}
\epsfxsize=15cm
\centerline{\epsfbox{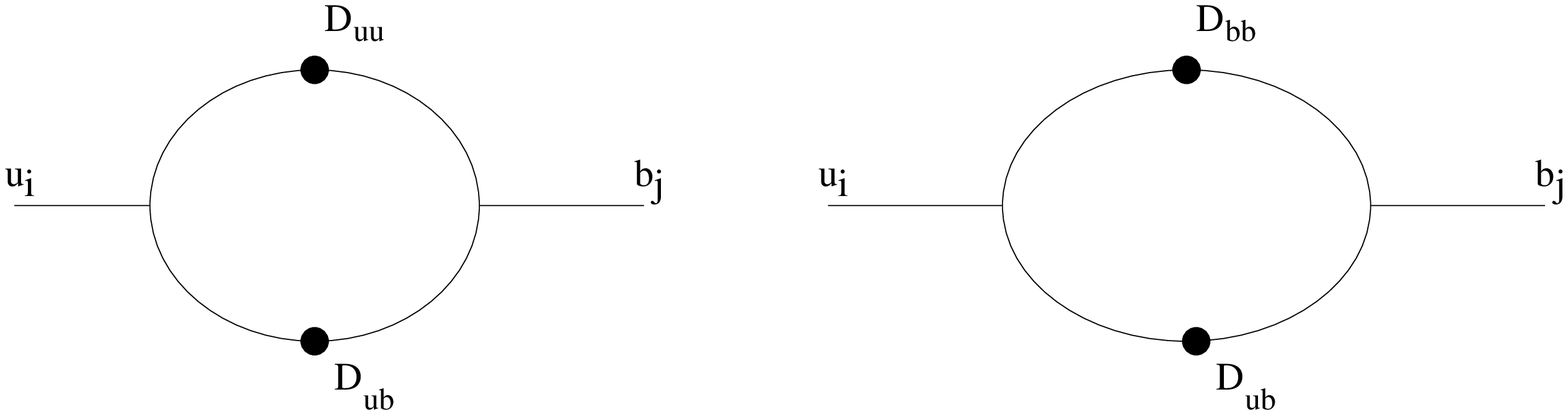}}
\caption{1-loop diagramatic correction to $D_{ub}$.}
\end{figure}
the form of the correction is 
\begin{equation}
{k_{1i}k_{2j}\over\ 16k^4}D_{uu}D_{ub}\int{d^3q\over\ (2\pi)^3}
{q^{\alpha}\epsilon_{lms}q_sq_lq_m\over\ q^4};
\end{equation}
however this correction vanishes for both $\alpha=0$ and 2 since the
integrand is odd in $q$. Hence there is no renormalisation of $D_{ub}$
and two-point structure functions Eqs.(4.22) and (4.23)
scale as $t^{-5/2}$ just as they do when $D_{ub}=0$.

\subsection{Higher-order correlation and structure functions:
anomalous temporal scaling}
We now  consider the $n$-point correlation functions with $n>2$. 
None of the three-point functions have any infrared divergence. 
Hence they exhibit simple scaling, i.e., $\,<u_i(x,t)^3>\sim t^{-9/2}$.
Let us consider the four-point functions namely, 
\begin{eqnarray}
&&<u_i({\bf x_1},t)u_j({\bf x_2},t)u_s({\bf x_3},t)u_l({\bf x_4},t)>,
\nonumber \\
&&<b_i({\bf x_1},t)b_j({\bf x_2},t)b_s({\bf x_3},t)b_l({\bf x_4},t)>,
\nonumber \\
&&<u_i({\bf x_1},t)u_j({\bf x_2},t)b_s({\bf x_3},t)b_l({\bf x_4},t)>,
\nonumber \\
&&<u_i({\bf x_1},t)u_j({\bf x_2},t)u_s({\bf x_3},t)b_l({\bf x_4},t)>.
\nonumber \\
\end{eqnarray}
At the 1-loop level, all of them they
will have finite parts (coming from initial auto correlations of velocity
and magnetic fields). However, the first three of them will have {\em new,
infrared-divergent}, one-loop contributions if $D_{ub}\neq 0$.
\begin{figure}[htb]
\epsfxsize=16cm
\centerline{\epsfbox{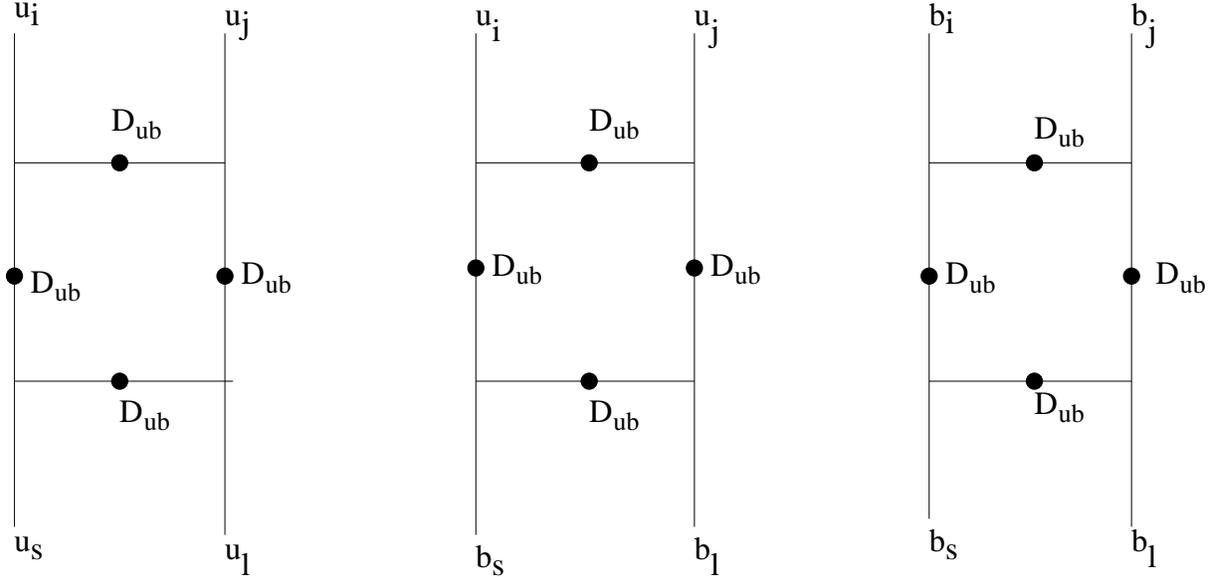}}
\caption{\small Divergent one-loop contributions to 
$\langle u_i({\bf x_1},t)u_j({\bf x_2},t)u_s({\bf x_3},t)u_l({\bf x_4},t)
\rangle$,
$\langle u_i({\bf x_1},t)u_j({\bf x_2},t)b_s({\bf x_3},t)b_l({\bf x_4},t)
\rangle$,
$\langle b_i({\bf x_1},t)b_j({\bf x_2},t)b_s({\bf x_3},t)b_l({\bf x_4},t)
\rangle$ respectively (see text).}
\label{Fig_4}
\end{figure}
The analytic structure of the divergent 1-loop integral is  illustrated for one
of these correlation functions in Fig.(\ref{Fig_4}) whose
contribution is
\begin{equation}
\langle u_i({\bf k_1})u_j({\bf k_2})u_s({\bf k_3})u_l({\bf k_4})\rangle 
\sim {k_{1i}k_{2j}k_{3s}k_{4l}\over\ k^8}\int {d^3p\over\ (2\pi)^3}
\epsilon_{acm}\epsilon_{cdn}\epsilon_{def}\epsilon_{eat}{p_mp_np_fp_t 
p^{\alpha} \over p^8}.
\label{4pt}
\end{equation}
Isotropy yields
\begin{equation}
p_mp_np_fp_t=Ap^4[\delta_{mn}\delta_{ft}+\delta_{mf}\delta_{nt}+
\delta_{mt}\delta_{fn}],
\label{delta}
\end{equation}
so summing over $(m,n)$ and $(f,t)$ separately we obtain
\begin{equation}
A={1\over\ d^2+2d}=1/15
\end{equation}
for $d=3$. Hence  naive power counting for
the integral in Eq.(\ref{4pt}) yields, for $\alpha=0$,
\begin{equation}
\sim{k_{1i}k_{2j}k_{3s}k_{4l}\over 15 k^8}D_{ub}^4
\int {d^3p p^4\over p^8}
\sim k_{1i}k_{2j}k_{3s}k_{4l}\Lambda^{-1}
\end{equation}
where $\Lambda$ is a momentum scale arising from the lower limit of the
integral. Hence it is linearly infrared (IR)
divergent for $\alpha=0$. For $\alpha=2$ there is no IR divergence as is the
case with $D_{ub}=0$. We analyse the case $\alpha=0$ in detail: There is
a disconnected diagram also that contributes to 
$\langle u_i({\bf x_1},t)u_j({\bf x_2},t)u_s({\bf x_3},t)u_l({\bf x_4},t)
\rangle$, which yields
\begin{equation}
D_{uuuu}\sim D_{uu}^2.
\end{equation}
This is finite in the $k\rightarrow 0$ limit, so the contribution from the 
connected part will dominate in the $k \rightarrow 0$ limit.
If we define the effective four-point
correlation to be $\propto D_{uuuu}\sim D^o_{uuuu}k^{-1}$ 
(with the choice $k=\sqrt{k_1^2+k_2^2+k_3^2+k_4^2}$), where $D^o_{uuuu}$ is just
a constant) then 
\begin{equation}
\langle u_i({\bf k_1},0)u_j({\bf k_2},0)u_s({\bf k_3},0)u_l({\bf k_4})
\rangle\;\sim
k_{1i}k_{2j}k_{3s}k_{4l}k^{-1}D^o_{uuuu} \delta({\bf k_1+k_2+k_3+k_4}).
\end{equation}
Hence one obtains
\begin{eqnarray}
<u_i ({\bf x},t)^4>&\sim&\int d^3q_1d^3q_2d^3q_3 exp(-k_1^2-k_2^2-k_3^2-(
k_1 +k_2+k_3)^2))t \nonumber \\
&&\times k_{1i}k_{2i}k_{3i}(k_{1i}+k_{2i}+k_{3i}) D^o_{uuuu} k^{-1}
\nonumber \\
&&\sim t^{-3d/2-2}t^{-1/2}\sim t^{-1/2}<u_i({\bf x},t)^4>|_{D_{ub}=0}.
\end{eqnarray}
Similarly $\langle b_i ({\bf x},t)^4\rangle\;
\sim t^{-1/2}\langle u_i({\bf x},t)^4\rangle |_{D_{ub}=0}, 
<u_i ({\bf x},t)^2b_i ({\bf x},t)^2>\sim t^{-1/2}<u_i ({\bf x},t)^2b_i ({\bf x},t)^2
>|_{D_{ub}=0}$.
The other four-point functions are (dimensionally) linearly divergent at the
1-loop level; however, the 1-loop integral vanishes becauses the integrand is odd
in $p$. Hence they will scale as $t^{-13/2}$.
It is now obvious that
\begin{eqnarray}
<[u_i({\bf x_1},t)-u_i({\bf x_2},t)]^4>&\sim& t^{-1/2}t^{-13/4}A_u
({{\bf x_1-x_2} \over t^{1/2}}) \nonumber \\
&\sim& t^{-1/2}<[u_i({\bf x_1},t)-u_i({\bf x_2},t)]^4>|_{D_{ub}=0}, \\
<[b_i({\bf x_1},t)-b_i({\bf x_2},t)]^4>&\sim& t^{-1/2}t^{-13/4}A_b
({{\bf x_1-x_2} \over t^{1/2}}) \nonumber \\
&\sim& t^{-1/2}<[u_i({\bf x_1},t)-u_i({\bf x_2},t)]^4>|_{D_{ub}=0}.
\end{eqnarray}
Hence we see that these structure functions exhibit anomalous dynamical scaling 
if $D_{ub}\neq 0$.  Similarly, it follows that the even-order exponents are
\begin{equation}
z_{2n}^u=z_{2n}^b={3\over 2}(2n-1)+n-z_{2n}^{ano},\;\;n\geq 2,
\end{equation}
where $z_{2n}^{ano}=(2n-3)/2$ is the {\em anomalous} part 
of the exponent $z_{2n}$. By contrast the odd order exponents
\begin{equation}
z_n^u=z_n^b=(n-1)3/2+n/2,\;\;{\rm for\; any\; odd}\; n,
\end{equation}
do not have an anomalous part.  
In Fig.(\ref{formula}) we present  plots of $z_n$ versus $n$
for both even and odd $n$. 
\begin{figure}
\centerline{
\epsfysize=9cm
{\epsfbox{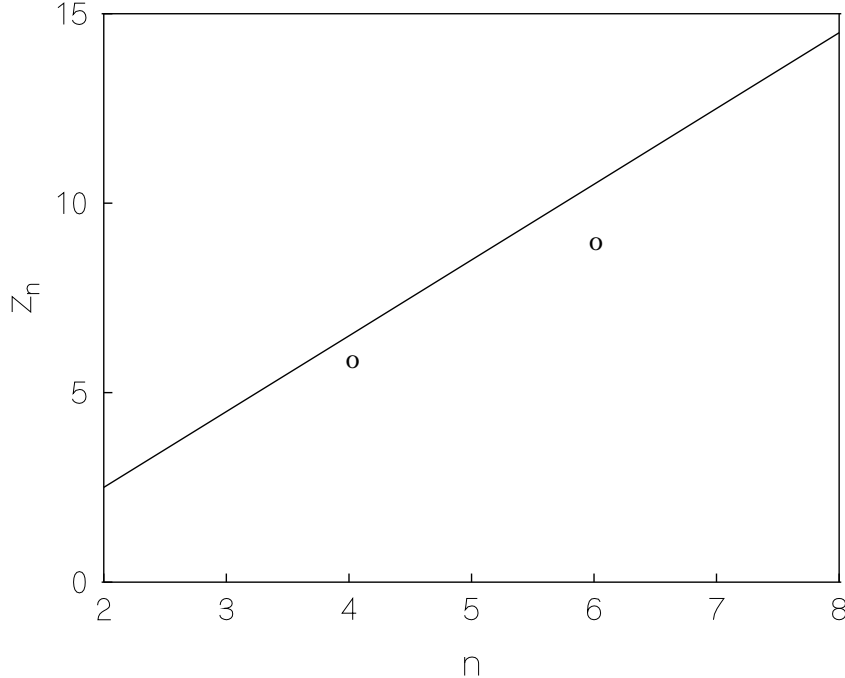}}
}
\caption{A plot showing the dependence of  
$z_{n}$ over $n$; the
line gives the exponents for the odd order structure functions for all values 
of $D_{ub}$ and for the even order structure functions 
when $D_{ub}=0$; we also show $z_4$ and $z_6$ in this plot (circles) when
$D_{ub}\neq 0$ (for $\alpha=0$).}
\label{formula}
\end{figure}

\section{Effects of initial conditions with singular amplitudes
for the variance}
In this Section we choose initial conditions of the following form:
\begin{eqnarray}
<u_i({\bf k},0)u_j({\bf k'},0)>&=&D_{uu}{k_ik_j\over k^{s+2}}\delta({\bf k+k'}),\\
<b_i({\bf k},0)b_j({\bf k'},0)>&=&D_{bb}{k_ik_j\over k^{s+2}}\delta({\bf k+k'}),
\\
{\rm and}\nonumber \\
<u_i({\bf k},0)b_j({\bf k'},0)>&=&iD_{ub}\epsilon_{ijp}{k_p\over k^{y+1}}
\delta({\bf k+k'}).
\end{eqnarray}
We choose\footnote {If $y>s$ then one-loop effective $D_{uu}$ and
$D_{bb}$ will scale as $k^{-y}$.}$s\geq y$.
The effective fluid viscosity $\tilde{\nu}$, and two-point correlations
$\tilde{D}_{uu}$ and $\tilde{D}_{uu}$ are given by
\begin{eqnarray}
{1\over\tilde{\nu}k^2}&=&{1\over\nu k^2}+{1\over\nu k^2}\int {d^3pp^{-s}p_ip_j
D_{uu}\over\nu[p^2+k^2-(k-p)^2]}[{1\over\nu[p^2+(k-p)^2-k^2]}(1-
{k^2\over\ p^2+(k-p)^2})\nonumber \\&-&
{1\over\ 2\nu p^2}(1-{k^2\over\ p^2}){\bf k.(k-p)}
+{1\over\nu k^2}\int {d^3pp^{-s}p_ip_jD_{bb}
\over\mu[p^2+k^2-(k-p)^2]}[{1\over\mu[p^2+(k-p)^2-k^2]}\nonumber \\&&(1-
{k^2\over\ p^2+(k-p)^2})-
{1\over\ 2\mu p^2}(1-{k^2\over\ p^2}){\bf k.(k-p)},\\
\tilde{D}_{uu}&=&D_{uu}+\Delta {D_{uu}^2\over\nu^2 k^{1-s}},\\
\tilde{D}_{bb}&=&D_{bb}+\Delta {D_{bb}^2\over\nu^2 k^{1-s}},\\
\tilde{D}_{ub}&=&D_{ub},
\end{eqnarray}
where $\Delta$ is just a numerical constant.
The self-consistent solution is given by 
$\nu,\mu\sim k^{1-s\over\ 2}$, with the 
long-wavelength behaviours of $D_{uu}, D_{bb}$
and $D_{ub}$ unchanged. Now we can easily calculate the
decay of the total velocity and magnetic energy, and cross helicity; these are:
\begin{eqnarray}
E^v(t)&\sim& t^{-{6-2s\over\ 2-s}},\\
E^b(t)&\sim& t^{-{6-2s\over\ 2-s}},
\end{eqnarray}
and
\begin{equation}
H^c(t)\sim t^{-{6-2y\over\ 2-s}}.
\end{equation}
Here $E^v=\int d^dk\,1/2\,<|v(k)^2>,\;E^b=\int d^dk\,1/2\,<|b(k)^2>,\;
H^c=\int d^dk\,|\bf v(k).b(k)|\,$ are total velocity-field and 
magnetic-field energies, and
total crosshelicity respectively. While calculating the temporal decay of these
quantities it has been implicitly assumed that the lower and
the upper limits of the momentum integrals can be extended upto 0 and 
$\infty$, respectively; if this not possible
 temporal behaviour cannot be obtained
by scaling (see next Section).

\subsection{Decay of fully developed MHD turbulence}
In fully developed $3d$MHD turbulence one finds, in the inertial range,
\begin{eqnarray}
\langle v_i({\bf k},t)v_j({\bf k'},t)\rangle &\sim & P_{ij}({\bf k})
k^{-11/3}\delta({\bf k+k'}),\\
\langle b_i({\bf k},t)b_j({\bf k'},t)\rangle &\sim & P_{ij}({\bf k})
k^{-11/3}\delta({\bf k+k'}).
\end{eqnarray}
Of course, a study of higher-order correlation functions reveals that the velocity
and magnetic- 
field distributions are not Gaussian \cite{abprl}. If, however, for the purpose
of analytical tractability, we assume that the initial distributions are Gaussian
with the variances as (for our Burges-type model)
\begin{eqnarray}
\langle u_i({\bf k},t=0)u_j({\bf k'},t=0)\rangle&=&D_{uu}k_ik_j|k|^{-11/3-2},\\
\langle b_i({\bf k},t=0)b_j({\bf k'},t=0)\rangle &=&D_{bb}k_ik_j|k|^{-11/3-2},
\end{eqnarray}
such that $|<u(k,t=0)u(-k,t=0)>|\sim |<b(k,t=0)b(-k,t=0)>|\sim
k^{-11/3}$ as observed in fully developed MHD turbulence,
we can calculate $z$ and effective two-point correlations as we did above in Sec
(4.2.3). For simplicity we give details only for our simple Burgers-type
model [Eqs.(\ref{mhd1}-\ref{mhd2})]. The effective $\nu$, $D_{uu}$ and $D_{bb}$ are given by
\begin{eqnarray}
&&{1\over\tilde{\nu}k^2}={1\over\nu k^2}+{D_{uu}\over\nu k^2}
\int d^d p{p_l p_m p^{-11/3-2}
{\bf k.(k-p)}\over\ 2\nu {\bf k.p}}[{1\over\ 2\nu p^2 -2\nu{\bf k.p}}
(1-{k^2\over\nu(p^2+({\bf k-p})^2}))\nonumber \\&-&{1\over\ 2\nu p^2}
(1-{k^2\over\ 2\nu p^2}]
+{D_{bb}\over\nu k^2}\int d^d p{p_l p_m p^{-11/3-2}
{\bf k.(k-p)}\over\ 2\mu {\bf k.p}}
[{1\over\ 2\mu p^2 }(1-{\nu k^2\over\ 2\mu p^2})- \nonumber \\ &&
{1\over\ 2\mu p^2 }(1-{\nu k^2\over\ 2\mu p^2})] \\
\tilde{D}_{uu}&=&D_{uu}+\delta {D_{uu}^2\over\nu^2 k^{8/3}}+
\delta {D_{bb}^2\over\mu^2 k^{8/3}}\\
\tilde{D}_{bb}&=&D_{bb}+2\delta{D_{bb}D_{uu}\over\mu^2 k^{8/3}}.
\end{eqnarray}
Here $\delta$ is a numerical constant. 
The self-consistent solutions of these equations are  given by $\nu,\mu \sim k^{-4/3}$ and 
$\tilde{D}_{uu}=\tilde{D}_{bb}$, implying that $z=2/3$; this is the same as
in the steady state. With this self-consistent solution, we see that
the 1-loop corrections to $D_{uu}$ and $D_{bb}$ are as singular as the 
bare quantities. Hence we ignore their renormalisations.
When $D_{ub}$ is non-zero, it will contribute to the 1-loop corrections
of $D_{uu}$ and $D_{bb}$ with the same low-wavenumber singularity.
$D_{ub}$ itself does not have any diagramatic correction, as in the 
case discussed in Section III.

The decay of two-point correlation functions will be of the form
\begin{equation}
\langle u_i({\bf x},t)^2\rangle \sim  \int d^3k\,\ exp(-2\nu k^{2/3} t)
k^{-22/3}
\end{equation}
It is easy to see that, because of the infrared divergence of the integral, 
the lower limit cannot be extended to 0, so it will
depend upon a lower cut-off (the inverse of which is the integral scale). 
Consequently the $t$-dependence cannot be scaled out as in
the previous case. This problem is reminiscent of the sweeping divergence
that appears in a perturbative DRG calculation of fully developed
turbulence or MHD turbulence \cite{epjb,4mou}. In fact this problem
begins when $<u_i({\bf k}, 0)u_j({\bf k'},0)>\sim k^{4-y-d}\; {\rm and}\;y=3$; 
then the above
integral becomes log-divergent. This type of initial state can be prepared
by randomly stirring the system with a stochastic force with a variance
$k^{4-y-d}$ with $y=3$. In a DRG analysis, sweeping divergences appear
when $y\geq3$ (but K41 scaling obtains for $y=4$)\cite{4mou}. 
Higher-order correlation functions have more severe divergences, hence
their decay also depend strongly upon the integral scale.

\section{Concluding remarks}
We have examined the decay of the total kinetic and magnetic
energies, and the crosshelicity in our shell model, These 
quantities decay with power laws in time. 
In order to understand this
behaviour analytically, we have employed a one-loop perturbation theory.
In the first part of our calculations, we have chosen 
random initial conditions with the statistical properties given in 
Eqs.(5),(6) and (10),(11) (with no crosscorrelations).
We have shown that such
initial conditions lead to power-law decay of temporal correlations
and appropriately defined structure functions. For initial correlations
analytic in the long-wavelength limit no higher order structure functions
show any anomalous scaling. However, on introducing 
crosscorrelations $<u_i({\bf k},0)b_j({\bf k'},0)>$ (Eq.4.43)
 higher-order  structure 
functions exhibit anomalous scaling. This highlights the importance of 
crosscorrreations in the statistical properties of MHD turbulence. We have also
extended our calculations to the case of initial correlations with 
singular variances and discussed the difficulties in
the decay of fully developed MHD turbulence. We have argued,
that these difficulties are related to the sweeping effect which arises in the
study of the stochsatically driven MHD turbulence (in the statistically 
steady state). It is also
worth noting that we have obtained similar temporal behaviour from both
$3d$MHD and our simplified Burgers-type model. 
It would be interesting to see if our results can be checked experimentally
and numerically.

\section{Acknowledgement}
The author wishes to thank Rahul Pandit, Sriram Ramaswamy and Jayanta K
Bhattacharjee for discussions. The author acknowledges CSIR (India) for
financial support.

\end{document}